 \documentclass[pre,preprint,showpacs,showkeys,amsfonts]{revtex4}

         \usepackage{latexsym}
         \usepackage{amssymb}
         \usepackage{graphicx}

          \newtheorem{lemma}{Lemma}[section]
           \newtheorem{theorem}{Theorem}[section]

\begin{document}
           \title
           {Zero-range potentials with Inner structure :\\ fitting parameters
            for  resonance  scattering}

 \author{Vladimir I. Kruglov}
 \email{vik@phy.auckland.ac.nz}
 \affiliation{Department of Physics, The University of Auckland,
Private Bag 92019, Auckland, New Zealand}
 
 \author{Boris S. Pavlov}
\email{pavlov@math.auckland.ac.nz}
\affiliation{Department of Mathematics,
University of Auckland, Private Bag 92019, Auckland, New Zealand}
 
	  \vskip 1truecm
           
           \thispagestyle{empty}
           \baselineskip=15pt
           \tolerance=600

\if01
\begin{center}
          {Vladimir I. Kruglov$^1$, Boris S. Pavlov$^2$}
\end{center}
{\it \,$^1$Department of Physics, The University of Auckland,
Private Bag 92019, Auckland, New Zealand. E-mail: {\tt
vik@phy.auckland.ac.nz} \\ $^2$Department of Mathematics,
University of Auckland, Private Bag 92019, Auckland, New Zealand.
E-mail: {\tt pavlov@math.auckland.ac.nz}}\\
 \fi
 
\vskip1cm
\begin{center}
{\it Dedicated to the centennial birthday of John von
Neumann,\\ Creator of the Operator-Extensions  Theory}\\
\vskip1cm

\end{center}

\begin{abstract}
 The solution of the  classical Fermi problem of low-energy neutron
scattering  by
nuclei, when the excitations of the  nuclei in scattering
processes are  taken into account, is found by the method of
zero-range potentials  with inner structure.
This  model is a generalization of the Fermi zero-range potential obtained by
adding a non-trivial inner Hamiltonian and  inner space with
indefinite metric. We propose a  general principle of {\it
analyticity  of the  Caley-transform  of the S-scattering matrix,}
written as a  function of
wave number  $k$. This permits us to evaluate all  parameters of  the  model,
including  the  indefinite  metric  tensor of  the  inner  space,
once the spectrum of the inner Hamiltonian, the scattering length and
the effective radius are chosen.
\end{abstract}
\maketitle 
     \vskip1cm
           \section{Introduction}

    Modelling the scattering problem of nucleons by nuclei using a
complex-valued potential leads to a basic physical model applicable
in the region of low energies \cite{Bohr,Gold}. This approach is called the
optical model and it is often used for the analysis of
experimental data. The parameters of the complex-valued potential are
fitted by solving the associated Schr\"{o}dinger equation and then
attempting to obtain the best
agreement with the experimental scattering cross-section. This
method requires laborious numerical calculations \cite{Bohr}.

In this paper we develop another model, called the zero-range
potential with inner structure, which serves as a generalization of the
Fermi zero-range potential model describing low-energy neutron
scattering by nuclei \cite{Fermi}.  The  Fermi zero-range potential
is generalization by adding a non-trivial inner Hamiltonian and
appropriate inner
space \cite{Zero_range} supplied with an indefinite metric. This
zero-range model with inner structure effectively takes into
account multiparticle interactions of the scattering particles which is
important for such problems as scattering of electrons by atoms and
scattering of nucleons by nuclei. The zero-range model with inner
structure is a solvable model which is an important feature of our approach.
This is the main advantage of our
model over the optical model. Moreover we do not introduce a
non-Hermitian Hamiltonian as in the optical model, hence the S-scattering
matrix of our model is unitary and has the correct placement of
poles and zeros in the complex plane of the wave number $k$.

It is known also \cite{Ohm} that the descriptions of the resonance processes
and collisions with redistribution are the most difficult problems in
the optical model. In the present work we consider the resonance
scattering in the s-channel where all results are given in explicit
analytical form. In this section we examine the main works
starting from the fundamental paper \cite{Fermi} which laid the foundation
for the
development of the zero-range model.
It is useful to study the Fermi model, assuming that the nucleus
potential is 3-D finite square well:
\begin{equation}
V(r) = \left\{
\begin{array}{ccc}
-V_{0},\,&\mbox{if}& r\leq  r_{0}\\
0 ,\,&\mbox{if}& r> r_{0}
\end{array} \right.
\label{i1}
\end{equation}
where $V_{0}>0$. Let us consider the bound state of the particle in
this well with the energy $E=-\frac{\hbar^{2}}{2\mu}\kappa^{2}<0$,
where $\mu =\frac{m_{1}m_{2}}{m_{1}+m_{2}}$ is the reduced mass of
scattering particles. Then the solution of the Schr\"odinger equation
decreasing as $r\rightarrow \infty$ is:
\begin{equation}
\psi(r) = \frac{C}{r}\exp (-\kappa r).
\label{i2}
\end{equation}
Defining the function $\chi(r)=r\psi(r)$ and using Eq.~(\ref{i2}) we
find the condition on the boundary of the 3-D square well :
\begin{equation}
\left(\frac{1}{\chi(r)}\frac{{\rm d}}{{\rm
d}r}\chi(r)\right)_{r=r_{0}} = -\kappa.
\label{i3}
\end{equation}
Let us consider the special case, i.e. the weakly bound state:
$|E|<<V_{0}$ or equivalently $\kappa r_{0}<<1$. This holds, for
example, in the case of deuteron and  negatively charged ions. If in this case
$\kappa$, or the energy of the bound state $E<0$, is fixed
and the inequalities $\kappa r_{0}<<1$ and $kr_{0}<<1$ hold, where
$k$ is the wave number, then the boundary condition (\ref{i3}) may be
reduced to the condition :
\begin{equation}
\left(\frac{1}{\chi(r)}\frac{{\rm d}}{{\rm
d}r}\chi(r)\right)_{r=0} = -\kappa.
\label{i4}
\end{equation}
This boundary condition was introduced by Fermi \cite{Fermi} in the scattering
problem of neutrons by nuclei and means that the wave function $\psi(r)$ is
   singular at $r=0$ and in the vicinity of the zero point ${\bf r}=0$ has
   the form:
\begin{equation}
\psi(r) \approx C\left(\frac{1}{r}-\kappa\right).
\label{i5}
\end{equation}
Introducing the Fermi s-wave scattering length $a\doteq \kappa^{-1}$
one can find that if $k\rightarrow 0$ then the total cross-section will be
$\sigma=4\pi a^{2}$. The boundary condition
(\ref{i4}) determines the so called the Fermi zero-range potential model
\cite{Fermi}.
It is important that the boundary condition (\ref{i4}) holds
in the low-energy limit $k\rightarrow 0$ for a deep well with arbitrary
form when the bound state with the energy
$E=-\frac{\hbar^{2}}{2\mu}\kappa^{2}<0$ is a weakly bound state.

We note that, in the general case,  the Laplace operator is symmetric on
the scattering waves of the Fermi zero-range potential. Really, if the
asymptotic of the scattering waves at the origin ${\bf r}=0$ is:
\begin{equation}
\psi_{i}(r) = \frac{A_{i}}{4\pi r}+B_{i}+o(1),
\label{i6}
\end{equation}
then the corresponding boundary form $J_{ik}$ for general smooth
elements $\Psi_{i}(r)$ possessing the above asymptotic at the origin
${\bf r}=0$ can be written
\begin{equation}
J_{ik}=\langle -\bigtriangleup \Psi_{i},\,\Psi_{k}\rangle -
    \langle  \Psi_{i}, -\bigtriangleup \Psi_{k}\rangle =
    \bar{A}_{i}B_{k}-A_{k}\bar{B}_{i},
\label{i7}
\end{equation}
where
\[
\langle \Psi_{i},\Psi_{k}\rangle = \int
\bar{\Psi}_{i}(r)\Psi_{k}(r){\rm d}^{3}{\bf r}.
\]
Due to Eq.~(\ref{i5}) the values $A_{i}$ and $B_{i}$ satisfy to the
{\it boundary condition}
\begin{equation}
\kappa A_{i}+4\pi B_{i}=0,
\label{i8}
\end{equation}
hence the boundary form (\ref{i9}) is zero: $J_{ik}\equiv 0$.

In the case of square well when the conditions $\kappa r_{0}<<1$ and
$kr_{0}<<1$ hold one can make a small correction to the Fermi
result :
\begin{equation}
\frac{1}{a} = \kappa -\frac{r_{0}}{2}\kappa^{2},
\label{i9}
\end{equation}
where $a^{-1}\doteq -(k{\rm cot}\delta(k))_{k=0}$ and the function
$k{\rm cot}\delta(k)$ has the form :
\begin{equation}
k{\rm cot}\delta(k) = -\frac{1}{a}+\frac{r_{0}}{2}k^{2}.
\label{i10}
\end{equation}
Here $\delta(k)$ is the scattering phase in the s-channell ($l=0$).
Eqs.~(\ref{i9}) and (\ref{i10}) are also correct in the low-energy
limit when the well has an arbitrary shape but the top bound state with the
energy $E=-\frac{\hbar^{2}}{2\mu}\kappa^{2}<0$ is the weakly bound state.
In this case the parameter $r_{0}$ is the effective radius of the theory
and is only in an approximate sense the well radius \cite{Ohm}.

A  single  standard  Fermi  zero-range  potential  defines  an
operator  with  only
trivial spectral  structure  (one negative eigenvalue). Still the
question remained until recently: is  it
possible  to  extend  the  construction  of  Fermi to  include
operators  with  rich spectrum? A version of this question
appeared in the paper \cite{Wiegner51} by Wiegner, where it was
noticed that for
compactly supported potentials $V(r)$ ($V(r)=0$ for $r>r_{0}$) the
scattering waves may be calculated for $r>r_{0}$ if we use an appropriate
energy-dependent boundary condition on the sphere $r=r_{0}$:
\begin{equation}
\left(\frac{\partial}{\partial
r}\chi(r,k)+\alpha(k)\chi(r,k)\right)_{r=r_{0}}=0,
\label{i11}
\end{equation}
where $\chi(r,k)=r\psi(r,k)$. Unfortunately this condition contains
the full scattering data, which correspond to compactely supported
potentials. Really, the solution of the Schr\"odinger equation leads
to the s-scattering waves at $r>r_{0}$ for compactely supported
potentials in the form :
\begin{equation}
\psi(r,k)=\frac{C}{r}{\rm sin}(kr+\delta(k)),
\label{i12}
\end{equation}
where $\delta(k)$ is the scattering phase in the s-channell $(l=0)$. Combining
Eqs.~(\ref{i11}) and (\ref{i12}) we find :
\begin{equation}
\alpha(k)=-k{\rm cot}(kr_{0}+\delta(k)).
\label{i13}
\end{equation}
Hence to calculate $\alpha(k)$ we should know $\delta(k)$ which is
equivalent to the solution of the Sch\"rodinger equation.
We note that the suggestion  of  Wigner  inspired
numerous  attempts to  construct the energy-dependent  potentials  and
appropriate boundary  conditions, see also \cite{Newton,PZ98}.

The authors of the paper \cite{BF61} suggested a new version
of the zero-range potential in  terms  of  von-Neumann
Operator-Extension techniques. All books  and  papers
on  zero-range  solvable models  in quantum
mechanics, see for  instance \cite{Albeverio, Kurasov}, are  based  on
this important paper \cite{BF61}.

Thereby Gelfand attracted \cite{SPN} attention  of mathematicians to the
necessity of
reconsidering the main results of von-Neumann's  Operator-Extension
theory for {\it general}  Hermitian operators in terms of
symplectic Hermitian forms similar to  the {\it boundary form} of
a differential operator. The above form (\ref{i7}), for instance
in $R_{1}$, is:
\begin{equation}
\label{i14}
\int_0^\infty
(-\bar{\psi}_{i}''(x)\psi_{k}(x)+\bar{\psi}_{i}(x)\psi_{k}''(x)){\rm
    d}x = (\bar{\psi}_{i}'(x)\psi_{k}(x)-\bar{\psi}_{i}(x)\psi_{k}'(x))_{x=0}.
\end{equation}
Realization of this plan would imply some sort of ``integration by
parts'' for abstract Hermitian operators, see  for  instance section 2 below.
    The first step in the development of symplectic techniques in
Operator-Extension theory was done in \cite{Zero_range} with the intention of
creating a quantum mechanical solvable  model with a  reasonably rich  and
algebraically  computable discrete spectrum and  resonances. This
program was  developed in a series of publications, see
for instance \cite{Gorbachuk,Extensions,Novikov,Kurasov}.
\par
In \cite{Zero_range} the ``zero-range potentials with inner
structure" were suggested as an application of the classical Krein
formula \cite{K,MN} to ``spectral modelling" of quantum systems.
This prospect was anticipated by M. Krein himself long before
\cite{VAD}. In particular the  operator extension approach permits one
to construct simple solvable models of quantum  systems with
resonance properties without solving sophisticated
boundary problems, but operating with  finite  matrices (see also
\cite{Adamyan,Exner,Ad_Pav} and more  references in
\cite{Kurasov}).
\par
    Unfortunately  zero-range solvable  models, in particular models with
inner structure, contain  a  large  number  of parameters  which
{\it have no  physical interpretation} and hence are not {\it fitting
parameters}. We propose a solution of this problem in this paper which has
application for low-energy
resonance scattering of neutron by nuclei. The generalization of the
Fermi zero-range potential is obtained by a symplectic version of the
operator-extention technique and assuming that the zero-range potential
has inner structure which arises by adding some ``inner'' Hamiltonian
and ``inner'' space $E$ with a special indefinite metric. Though our
treatment is based on Operator-Extention techniques we are able
to define generalized zero-range potentials by {\it special boundary
conditions} at the origin ${\bf r}=0$, which is a generalization of
the Fermi condition (\ref{i4}) :
\begin{equation}
\left(\frac {1}{\chi({\bf r},k)}\frac{\partial}{\partial r}\chi({\bf
r},k)\right)_{r=0} = k{\rm cot}\delta(k).
\label{i15}
\end{equation}
This condition can be found formally if we suppose that the scattering
wave has the form (\ref{i12}) at the origin ${\bf r}=0$---which is not
trivial because, for conventional potentials, it is not generally true. It is
important to understand that this boundary condition does not define
the particular
zero-range potential but it defines a whole class of zero-range
potentials.
In section 5 we show that the zero-range potentials derived in our
paper do in fact satisfy the boundary condition (\ref{i15}).

We note that the boundary condition (\ref{i15}) follows from the
expantion of the wave function $\psi({\bf r},k)$ at the origin ${\bf
r}=0$ as:
\begin{equation}
\psi({\bf r},k) = C{\rm sin}\delta(k)\left(\frac{1}{r}+ k{\rm
cot}\delta(k)+\ldots\right).
\label{i16}
\end{equation}

Another, most important, point of our treatment is the {\it principle
of analyticity} of the function:
\begin{equation}
k{\rm cot}\delta(k) = -\frac{1}{a}+ \frac{r_{0}}{2}k^{2}+
\ldots=\sum_{n=0}^{\infty}g_{n}k^{2n}.
\label{i17}
\end{equation}

Thus we require that this function is an {\it entire function} of the
complex variable $k$ which permits us to evaluate all the extra parameters of
the zero-range model with inner structure.

In the  next section we  develop  symplectic techniques  for  the  general
extension procedure  of  symmetric  operators  in a Pontryagin space
    (this may be thought of as an abstract  version of  integration by  parts).
In the  third  section we  calculate  the  S-scattering  matrix based
on a variant  of  the  Krein formula. Then in section 4 we study a
special class of  zero-range  models of  scattering  systems  for  which
the scattering  matrices fulfil  the  analyticity  condition
formulated  for  the function $k{\rm cot}\delta(k)$. It  appears that
for the models of this  class, {\it  if the scattering length,
effective radius and the eigenvalues of the
    inner Hamiltonian are fixed, then all other parameters of the model,
    in particular  the boundary parameters and the deficiency vector
     in the finite-dimensional case
dim$\, E = N$ may be calculated directly}. Moreover, the type of the
indefinite metric in the inner space $E$ is also pre-defined  by
the above condition of analyticity. \\
Then in section 5  we  apply  the  results to a typical problem of
resonance  s-scattering. In Appendix A we give a
straightforward derivation  of  the  basic  Krein formula.
One usually assumes that the simplest
zero-range potential is the singular potential proportional to the Dirac
$\delta$-function, which is often used in various physics problems.
Consequently, we provide additional on this case in Appendix B.

\vskip1cm
\section{Symplectic version of the Operator-Extension techniques}

\noindent At first we define the dimensionless coordinates $x_{i}$
in $R_{3}$ space  by the equation ${\bf x}=k_{0}{\bf r}$  assuming
that $k_{0}$ is some characteristic wave number. We also assume that
the wave functions $u({\bf x})$ are dimensionless then the
Schr\"odinger equation for scattering problem can be written:
\begin{equation}
\left(\triangle +\lambda -\tilde{V}({\bf x})\right)u({\bf x}) = 0,
\label{s1}
\end{equation}
where $\lambda=E/E_{0}$ is a dimensionless energy with
$E=\frac{\hbar^{2}k^{2}}{2\mu}$ and $\tilde{V}({\bf x})$ is the
dimensionless potential
\begin{equation}
\tilde{V}({\bf x}) = E_{0}^{-1}V({\bf
r}),~~~E_{0}=\frac{\hbar^{2}k_{0}^{2}}{2\mu}.
\label{s2}
\end{equation}
Here $\mu = m_{1}m_{2}/(m_{1}+m_{2})$ is the reduced mass of the
scattering particles.
Below we will use conventional notations
\begin{equation}
x:={\bf x}=k_{0}{\bf r},~~~\lambda:=\frac{k^{2}}{k_{0}^{2}},
\label{s3}
\end{equation}
then the equation for the Greens function of free evolution is:
\begin{equation}
-(\triangle +\lambda)G_{\lambda}(x,x') = \delta^{(3)}(x-x'),
\label{s4}
\end{equation}
and the appropriate Greens function for outgoing waves has the form:
\begin{equation}
G_{\lambda}(x,x') = -(\triangle+\lambda)^{-1}\delta^{(3)}(x-x') =
\frac{e^{i\sqrt{\lambda}|x-x'|}}{4\pi|x-x'|}.
\label{s5}
\end{equation}

We  develop here  the  symplectic  technique  for the
zero-range potential of a pair of quantum  particles in
three-dimensional space, assuming that  in the course of  interaction
the pair may enter  an  ``inner''  space $E$ supplied with an {\it indefinite
metric}. This interection is determined by an inner Hamiltonian  ${\bf B}$,
which is
a Hermitian operator with respect to the properly defined
dot-product [.\,,\,.] in $E$. The two-body Hamiltonian
in the outer  space,  with respect to  coordinates defined in terms of
the  center of mass of  the  pair, may be given by a three-dimensional
self-adjoint
``non -perturbed'' Schr\"{o}dinger operator in the ``outer'' space
$L_2 (R_3)$ with the conventional dot-product $\langle u, \alpha
v\rangle = \alpha \langle u, v\rangle = \langle \bar{\alpha}u,
v\rangle$:
\begin{equation}
\label{nonpert} lu = -\bigtriangleup u,
\end{equation}
defined  on  the  Sobolev class $W_2^2(R_3)$ of  all
square-integrable functions  $u$ possessing  square integrable
second-order derivatives. In the course of  our  analysis the
spectral parameter may take negative and complex  values. The
operator $l = - \bigtriangleup$  is defined with respect to  the
dimensionless space variable $x$.
Other variables $\tilde {x},\tilde {y}$ will  be  used
for  vectors in the  inner  space $E$ of  the  model, see
below.
\par
Following  \cite{BF61} we  restrict  the operator $l$ to  the
domain consisting of all smooth functions  vanishing near
   $x=0$. The closure $l_0$ of the restricted
operator is defined on the domain $D_0$ which  consists  of all
$W^2_2$-functions vanishing just at the origin $x=0$. This  operator is
symmetric  and  has deficiency indices  $(1,1)$:
\[
\displaystyle \overline{(l_0 - i I){D}_0} = L_2(R_3)\ominus
N_i,\,\, \overline{(l_0 + i I){D}_0} = L_2(R_3)\ominus N_{-i},
\]
dim $N_i = $ dim $N_{-i}= 1$. The  adjoint  operator $l_0^{+}$ is
defined, due  to the von-Neumann theorem, see \cite{Akhiezer}, by the
same  differential  expression (\ref{nonpert}) on the  domain  $
D_0^+ = D_0 + N_i + N_{-i}$, where the  one-dimensional  subspaces
$ N_i ,  N_{-i}$ are  spanned by  the corresponding non-perturbed Green
functions $G_{\lambda}(x,0)$ at ${\rm Im} \sqrt{\lambda}>0$ and $\lambda =
\pm i $~:
\[
N_i = \left\{G_{-i}(*,0)\right\},\,\,N_{-i} =
\left\{G_{i}(*,0)\right\},
\]
\begin{equation}
\label{defect} G_{-i} (x,0)=  \frac{e^{i \sqrt{-i}
|x|}}{4\pi|x|},\,G_{i} (x,0) = \frac{e^{i \sqrt{i} |x|}}{4\pi|x|}
= \frac{l + iI}{l-iI} G_{-i} (x,0),
\end{equation}
where  we define  the branch of the square  root by the condition
${\rm Im} \sqrt{\lambda} > 0.$
\par
The  above description of  the  domain of the adjoint operator
based on the von-Neumann formulae was used  in \cite{Akhiezer,BF61}.
The other description of  the  domain  of  the adjoint operator
as the set  of  singular  elements  described  above
    (\ref{i6}) was
probably invented by  Fermi \cite{Fermi} and  is now commonly
used in the physical  literature, see  for instance
\cite{Demkov,Albeverio}. Each element $u$ from the domain of the
adjoint operator is characterized  by  the  {\it asymptotic
boundary  values } $A,\,\,B$ for $|x|\to 0$:
\begin{equation}
\label{fermivar} u (x) = \frac{A^u}{4\pi |x|} + B^u + o(1).
\end{equation}
It  is  well known that these representations  are equivalent.
Nevertheless we have a
good reason, see  Lemma 2.2, to derive this fact  from the decomposition
of elements of the defect $N_i + N_{-i}$ with respect  to  the
{\it symplectic basis}
\[
W_+ (x)= \frac{1}{2}\left[G_{-i}(x,0) + G_{i}(x,0) \right] =
\frac{l}{l - iI}G_{-i}(x,0),
\]
\[
W_{-} (x)= \frac{1}{2i}\left[ G_{-i}(x,0)  -  G_i (x,0) \right ]=
- \,\,\, \frac{I}{l - iI}G_{-i}(x,0).
\]
We  will present the von-Neumann  formula for  elements of the
domain of the adjoint operator in terms of a  decomposition  with respect
to the  basis $W_{\pm}$~:
\begin{equation}
\label{sympl3} u = u_0 + \eta^u_+ W_+ + \eta^u_- W_-,
\end{equation}
where $u_0$ is  an  element from the  domain of  the closure $l_0$
of the restricted operator and $\eta^u_{\pm}$ are
complex  coefficients. The deficiency elements $G_{\pm i}(*,0)$
are the eigenvectors of the adjoint  operator with eigenvalues
$\mp i $ respectively. Then  the elements $W_{\pm}$ are
transformed by  the  adjoint  operator as:
\begin{equation}
\label{Wpm}
     l_0^+ W_+ = W_{-},\,\,
      l_0^+ W_{-}= - W_{+}.
\end{equation}
\begin{lemma}{ The  boundary form
\[
\langle l_0^+ u,\,\, v \rangle - \langle u,\,\, l_0^+ v \rangle :=
J_{l}(u,v),
\]
of the adjoint operator $l_0^+$ depends only on the terms $u_d, \, v_d
$ of the elements $u,\, v$ in the  defect. It is an Hermitian
symplectic form in the variables $\eta_{\pm}$ and may  be presented as
\[
     J_{l}(u,v) =  J_{l}(u_d,v_d)=  \langle l_0^+u_d
,\,\, v_d \rangle - \langle u_d,\,\, l_0^+ v_d \rangle =
\]
\begin{equation}
\label{intbp} \frac{1}{4\pi\sqrt{2}} \left(\eta^u_+ \bar{\eta}^v_-  -  \eta^u_-
\bar{\eta}^v_+\right).
\end{equation}
The symplectic variables $\eta_{\pm}$ are
connected with the  asymptotic boundary  values $A,\, B$ via the
transformation:
\begin{equation}
\label{J-unit}
\left(
\begin{array}{c}
\eta_+\\
\eta_-
\end{array}
\right) = \left(
\begin{array}{cc}
1&0\\
-1& -4\pi\sqrt{2}
\end{array}
\right) \left(
\begin{array}{c}
A\\
B
\end{array}
\right).
\end{equation}
Then  the Hermitian symplectic form is given by $J_l (u,v) =
\left(B^u \bar{A}^v -
A^u \bar{B}^v\right)$. }
\end{lemma}

{\bf Remark}\,\, The  above representation  of the  ``outer''
boundary form in terms of the symplectic variables  $A,B$ is
usually obtained, see for  instance \cite{Fermi,BF61, Demkov,
Albeverio} via straightforward integration by parts. We use the
construction suggested in the above lemma to introduce new symplectic
variables which are useful in the symplectic version of operator-extension
theory for general ``abstract" operators, see below.

     {\it Proof}\,\,\, of  the  formula  (\ref{intbp}) may be  obtained
via direct
application  of  the  adjoint  operator  to  the above version
(\ref{sympl3})   of  the  von-Neumann  decomposition of elements
from $D_0^+$:
     \[
        l_0^+ u =  l_0 u_0 +  \eta^u_+ W_{-} -
          \eta^u_- W_+,
     \]
in the  boundary  form  of  the  adjoint  operator:
\begin{equation}
\label{bforml}
     \langle l_0^+
u,\,\, v \rangle - \langle u,\,\, l_0^+ v \rangle := J_{l}(u,v).
\end{equation}
One can easily  see that  the above  ``outer''
boundary form depends only on the parts of the elements $u,\,\,v$ in
the  sum of the deficiency subspaces (on the ``defect''), $u_d =
\eta^u_+ W_+ + \eta^u_- W_-,\,\, v_d = \eta^v_+ W_+ + \eta^v_-
W_-$:
\[
J_{l}(u,v) = J_{l}(u_d,v_d).
\]
Now the  boundary form  may be  calculated (using (\ref{Wpm})) :
\[
\langle l_0^+u_d ,\,\, v_d \rangle - \langle u_d,\,\, l_0^+ v_d
\rangle = \left(\eta^u_+ \bar{\eta}^v_-  - \eta^u_-
\bar{\eta}^v_+\right) \int_{R_3} |G_{i}(\xi,0)|^2 d^3\xi.
\]
The integral is evaluated as $1/(4\pi \sqrt{2})$ giving us
the desired formula (\ref{intbp}). The second formula is based
on the calculation of the asymptotic behavior of the non-perturbed
Greens function as $x \to 0$ :
\[
W_+ (x)= \frac{1}{4\pi |x|} - \frac{1}{4\pi\sqrt{2}}+ o(1),\,\,
W_- (x)= - \frac{1}{4\pi\sqrt{2}}+ o(1).
\]
Hence for $x \to 0$ we have the asymptotic behavior:
\[
\eta_+ W_+ (x) + \eta_- W_- (x) =  \frac{\eta_+}{4\pi |x|} -
    \frac{\eta_- + \eta_+ }{4\pi \sqrt{2}}  + o(1).
\]
The second  announced formula  follows immediately from the  above
asymptotics.

$\Box$

Self-adjoint  extensions  of  $l_0$ were  obtained  in
\cite{Fermi,BF61} by requiring that the asymptotic boundary values
$A,\,B$ satisfy Fermi boundary conditions  $ B = \gamma A $ parametrized
by a  real $\gamma = \bar{\gamma} $. In \cite{Zero_range} an
orthogonal sum of two operators  $l_0 \oplus {\bf B}_0 $ with an
abstract operator ${\bf B}$ was restricted and then  extended based on
an abstract version of the above symplectic ``integration by
parts'' (\ref{intbp})  in a Hilbert space $ E$ with positive
metric, ${\bf B} : E \to E$. We will now develop the symplectic technique
for a $g$-Hermitian operator ${\bf B}$ in a Hilbert Space space {\it
with an indefinite metric}.
\par
Assume that  $P_{\pm}$ are two complementary orthogonal
projections in a finite-dimensional Hilbert space $E$,\, dim$E = N $
with conventional dot-product $\langle*,*\rangle$  and consider the
operator $g = P_+ - P_-$ defining an indefinite dot product ($g$-dot
product):
\[
\langle g\tilde {x},\,\tilde {y} \rangle =
[\tilde {x},\tilde {y}]= \langle P_+ \tilde {x},P_+ \tilde {y} \rangle -
\langle P_- \tilde {x},P_- \tilde {y} \rangle.
\]
Here $ [\bar{\alpha} \tilde {x},\tilde {y}] = {\alpha}[\tilde {x},\tilde {y}]=
[\tilde {x},\alpha \tilde {y}]$. We denote by $E_g$ the space
$E$ supplied with the $g$-dot product. It is a
     Pontryagin space with indefinite metric $g$
     if dim $P_-$ is finite, see for  instance \cite{Krein-Langer,
Shondin2}.
We develop below a symplectic version of the operator-extension
technique in  the Pontryagin space which  differs  from the  above
techniques in  $L_2 (R_3)$ by minor details only.
\par
     A bounded or densely defined closed operator ${\bf B}$ is called
$g$-symmetric (symmetric with respect to a $g$-dot product) if $[{\bf
B}\tilde {x},\tilde {y}]=[\tilde {x},{\bf B}\tilde {y}]$ at least on the
domain of ${\bf B}$ in $E$.
This condition is equivalent to the condition $g {\bf B} \subset
{\bf B}^+ g$, where ${\bf B}^+$ is the adjoint operator with
respect to the conventional dot-product $\langle*,*\rangle$ in
$E$. Without loss of generality we may assume that the operator
${\bf B}-iI$ is invertible and the inverse is bounded.
\par
Eventually we will apply the  symplectic scheme to bounded operators
which are symmetric with
respect to the conventional dot-product in $E$ and commute with
$g$ in the finite-dimensional space $E$ (dim$E = N$). We  assume  that
the spectrum of the operator ${\bf B}$ consists of a  finite
number of simple  real eigenvalues  $\lambda_s$ :
\[
\left({\bf B} - \lambda_s \right)e_s= 0.
\]
One may choose a normalized {\it generating} vector \footnote{
``Generating" means: non-orthogonal to all eigenvectors $e_s$ of
the operator ${\bf B}$, $[e,e_s]\neq 0 ,\, s = 1,2,\dots N$. }
$e\in E,~~[e,e]= 1$ such that the vectors $e$ and $e' = \frac{{\bf
{\bf B}}+iI}{{\bf B}-iI} e$ are connected by the above
$g$-isometry $\frac{{\bf B}+iI}{{\bf B}-iI} e$ and form a
linearly-independent pair (${\bf B} e\neq 0$). We define the domain
$D_{{\bf B}_0}$ of the restricted operator as $D_{{\bf {\bf
B}}_0}=\left[{\bf B} - iI\right]^{-1} \left\{ E \ominus_g \left\{e
\right\}\right\}$ and set ${\bf B}_0 = {\bf B} \big|_{D_{{\bf
B}_0}}$. Here the orthogonal difference with respect to the $g$-dot
product is denoted by $ \ominus_g $. The vector $e$ plays, in this
construction, the role of the deficiency vector of the restricted
operator at  the spectral point $i$:
      \[
[({\bf B} - iI)D_{{\bf B}_0}, e]= 0,
      \]
and  the  vector $e'$ $([e',e']= [e,e])$ plays a role of  the
deficiency vector at  the  spectral point $- i$:
     \[
[({\bf B} + iI)D_{{\bf B}_0}, e']=0.
     \]
     The  sum $M$ of  two one-dimensional subspaces $M_i,\, M_{-i}$
spanned  by
     the  vectors  $e,\, e'$ respectively is called  the  {\it defect} of
the  operator ${\bf B}_0$. At  any  other  two  complex
conjugated regular points $\lambda,\, \bar{\lambda}$ of  the
operator ${\bf B}$ the  deficiency  subspaces
$N_{\lambda},\,N_{\bar{\lambda}}$ are one-dimensional. The
corresponding deficiency vectors  are calculated  as
\[
e_{\lambda}= \frac{{\bf B} + iI}{{\bf B} - \bar{\lambda} I} e,\,\,
e_{\bar{\lambda}}= \frac{{\bf B} + iI}{{\bf B} - {\lambda} I} e,
\]
and  the  defect  $M_{\lambda} + M_{\bar{\lambda}} $ is
two-dimensional. Let us define  the  $g$-adjoint operator ${\bf B}_0^+$
by the formula
\[
[{\bf B}_0 \tilde {x},\tilde {y}]= [\tilde {x},{\bf B}_0^+ \tilde
{y}],
\]
on the elements  $\tilde {y}$ for  which the $g$-dot product in the
left hand side
may be  continued  onto the  whole  space $E$ as an ``anti-linear"
functional \footnote{Recall  that we are using the ``physical''
notations for  the  dot-product $[\bar{\alpha}\tilde {x},\,\tilde {y}]=
\alpha[\tilde {x},\,\tilde {y}]= [\tilde {x},\, \alpha \tilde {y}] $} of
$\tilde {x}$.
For the densely-defined operator ${\bf B}$ and $e$ chosen from the complement
of the  domain $D_{\rm B}$ this condition implies
\begin{equation}
\label{adjvec}
     ({\bf B}^+_0 + iI)e = 0,\,\, ({\bf B}^+_0 -
iI)e' =
     0.
\end{equation}
For a bounded operator ${\bf B}$ we just  define the  adjoint operator
on the  defect  by  the above  formula (\ref{adjvec}). Then for
any  complex  value  of  the  spectral parameter  the  deficiency
vector  is  the  eigenvector of the  adjoint operator at  the
adjoint  spectral point:
\[
\left({\bf B}_0^+ - \bar{\lambda} I\right) e_{\lambda} = 0,\,\,
\left({\bf B}_0^+ - \lambda I\right) e_{\bar{\lambda}}=0 .
\]
This is similar to  the above fact for
Laplacian :
\[
     l_0^+ G_{\lambda}= \lambda  G_{\lambda},
\]
{\it on the  space  of  square integrable  functions} $L_2 (R_3)$.
\par
     An analog of the von-Neumann representation of the
domain of the adjoint operator remains true in a Ponryagin space,
see \cite{Krein-Langer}. So,  for elements from
the domain of the  adjoint operator we have the representation $\tilde {x} =
\tilde {x}_0 +
\alpha e + \beta e',\,\, \tilde {x}_0 \in D_0$ with
\[
{\bf B}_0^+ (\tilde {x}_0 + \alpha e + \beta e') = {\bf B} \tilde {x}_0 -i
\alpha
e + i \beta e'.
\]
Note  that  in the case when the  original operator ${\bf B}$ commutes with
$g$ and  is self-adjoint with respect  to  the  conventional
dot-product, the operator $\frac{{\bf B} + iI}{{\bf B} - iI}$ is
isometric both in the conventional and $g$-metric.
\par
     Consider a new  basis for the defect $M  = M_i  + M_{-i}$ of the
      operator ${\bf B}_0$, which is similar
to the  above  basis consisting of  linear  combinations of  Greens-functions:
\[
w_+ = \frac{e + e'}{2} = \frac{{\bf B}}{{\bf B} - iI} e ,\,\,w_- =
\frac{e - e'}{2i} = \frac{- I}{{\bf B} - iI} e.
\]
We  see  from (\ref{adjvec}) that
\[
{\bf B}^+_0 w_+ =  w_-,\,\, {\bf B}^+_0 w_- = -  w_+.
\]
     In the symplectic version of operator-extension theory
we use the new  basis to represent  elements from the domain of
the adjoint  operator via  new dimensionless symplectic
variables $\xi_{\pm}$ which play a role analogous to  the above
pair $\eta_{\pm}$:
\[
\tilde {x} = \tilde {x}_0 + \xi^x_+ w_+ + \xi^x_- w_-.
\]
\begin{lemma}
{The  boundary  form of  the  adjoint  operator may  be
given by  an Hermitian symplectic form in terms of the variables
$\xi_{\pm}$:}\end{lemma}
\begin{equation}
K(\tilde {x},\tilde {y})=\left[{\bf B}^+ \tilde {x},\,\tilde {y}\right]-
\left[\tilde {x},\, {\bf B}^+ \tilde {y}\right]=
\xi^x_+ \bar{\xi}^y_- -
\xi^x_- {\bar{\xi}}^{y}_{+}.
\label{boundf}
\end{equation}
{\it Proof}  exactly   follows the pattern of  the previous lemma
concerning the representation of the  boundary form in the ``outer''
space $L_2 (R_3)$. Really, the value  of  the boundary  form
depends  only  on the parts of  the vectors  $\tilde {x},\tilde {y}$
in the defect,
\[
K(\tilde {x},\tilde {y})=
\]
\[
[{\bf B}_0^+ (\xi^x_+ w_+ + \xi^x_- w_-),\, (\xi^y_+ w_+ + \xi^y_-
w_-)] - [(\xi^x_+ w_+ + \xi^x_- w_-),\, {\bf B}_0^+ (\xi^y_+ w_+ +
\xi^y_- w_-)]=
\]
\[
[(\xi^x_+ w_- - \xi^x_- w_+),\,
(\xi^y_+ w_+ + \xi^y_- w_-)] -
[(\xi^x_+ w_+ + \xi^x_- w_-),\, (\xi^y_+ w_-
- \xi^y_- w_+)]=
\]
\[
\xi_+^x \bar{\xi}^y_-([w_-,w_-]+[w_+,w_+]) - \xi^x_-
\bar{\xi}^y_+ ([w_-,w_-]+[w_+,w_+])=
\xi^x_+ \bar{\xi}^y_- -
    \xi^x_-{\bar{\xi}}^{y}_{+},
\]
since the  coefficients $([w_-,w_+] - [w_+,w_-])$ in front of
$\xi^x_{\pm} \bar{\xi}^y_{\pm}$ are  equal to  zero. Thus we have :
\[
w_+ = \frac{{\bf B}}{{\bf B} - i I}e,\,\, w_- = \frac{-I}{{\bf B}
- i
    I} e,
\]
which implies $([w_-,w_-]+[w_+,w_+])= 1$ and $([w_-,w_+] -
[w_+,w_-]) = 0$ and  the  announced abstract integration by parts formula
(\ref{boundf}).

$\Box$

Consider now the  orthogonal sum  of  operators $l\oplus {\bf B}$
in the Pontryagin space  $L_2 (R_3)\oplus E_g $ with elements $U =
(u,\,\tilde {x}),\, u\in L_2(R), \,\tilde {x}\in E_g $. Restricting both
operators
as above we obtain the symmetric operator $l_0 \oplus {\bf B}_0 $
with defect $N\oplus M$ and deficiency index $(2,2)$. We define
the adjoint operator as an orthogonal sum of adjoints $l_0^+
\oplus{\bf B}_0^+$ and calculate the boundary form of  it as a sum
of boundary forms of the  outer and  inner  components:
\[
{\cal J}(U,V)= J(u,v)+ K (\tilde {x},\tilde {y}) = (B^u \bar{A}^v - A^u
\bar{B}^v )+(\xi^x_+ \bar{\xi}^y_- - \xi^x_- \bar{\xi}^y_+),
\]
where we  choose the  symplectic variables  $A,B$ for the ``outer"
boundary form. The boundary  form  ${\cal J}(U,V)$  of  the
orthogonal sum  of operators $l_0^+ \oplus {\bf B}_0^+$ in the
Pontryagin  space $L_2 \oplus_g E_g$  is a symplectic Hermitian
form. Hermitian extensions  of  the  operator $l_0 \oplus {\bf
B}_0$ (or Hermitian  restrictions of the  operator $l_0^+ \oplus
{\bf B}_0^+$) may be constructed  via  imposing  boundary
conditions  on symplectic variables $A,B,\xi_{\pm}$ such  that the
total boundary form vanishes on  the corresponding  planes. For
instance the condition defined by a Hermitian matrix is
\[
\Gamma = \Gamma^+ = \left(
\begin{array}{cc}
\gamma_{00} & \gamma_{01}\\
\gamma_{10} & \gamma_{11}
\end{array}
\right) ,
\]
\begin{equation}
\label{bcond}
\left(
\begin{array}{c}
B \\
-\xi_-
\end{array}
\right) =
\left(
\begin{array}{cc}
\gamma_{00}  & \gamma_{01} \\
\gamma_{10} & \gamma_{10}
\end{array}
\right)
\left(
\begin{array}{c}
A \\
\xi_+
\end{array}
\right).
\end{equation}
The  planes  in  the  four-dimensional space of complex symplectic
variables $A,B,\xi_{\pm}$ where these boundary  conditions  are
fulfilled and hence the  boundary  form vanishes, are called {\it
Lagrangian planes} of the symplectic form ${\cal J}(U,V)$. We
assume here that the  elements $\gamma_{ik}$ have  numerical
values.
\par
All Lagrangian planes of the  above form  may be constructed
either with a help of a Hermitian matrix $\Gamma$, as above, or
obtained from the constructed plane by a proper $\cal J$-unitary
transformation, for instance:
\[
A \longrightarrow A' ,\,\,B \longrightarrow B' ,
\]
\[
\xi_- \longrightarrow \xi'_-,\,\, \xi_+ \longrightarrow  \xi'_+
\]
which  leaves  the  form ${\cal J}(U,V)$ unchanged. One  may  show
that all  Lagrangian planes  of the  form  ${\cal J}(U,V)$ may  be
obtained via  the composition of  Hermitian  transformations
(\ref{bcond}) and the fundamental ${\cal J}$-unitary  transformation
$ A' = A,\,\, B' =  B ,\, \xi'_- = \xi_+,\,\, \xi'_+ = - \xi_-$.
\par
The  matrix elements  $\gamma_{ik}$ are  called  ``the boundary
parameters ". They do  not have  any  straightforward  physical
interpretation. One of the most important problems  of the theory of
solvable  models
based  on  operator-extension  methods is  the  problem of {\it
fitting} these parameters in a physically  meaningful way. We solve this
problem for the general case in section 5.
\vskip1cm

\section{Krein formula and the S-scattering matrix}

The  expression for the non-stationary S-scattering
matrix  obtained  as  a  product  of  wave-operators  may  be
derived \cite{Ad_Pav} from  the corresponding  Krein
formula \cite{K,MN}. This formula gives a description of all
resolvents of an extended operator in terms of the boundary
conditions (\ref{bcond}) and some functional parameter---the so called
Krein's $Q$-function, see \cite{Akhiezer}. In this paper we
focus on the derivation  of an  expression for the {\it stationary
S-scattering matrix}, just fitting the properly constructed  ansatz
for  scattered  waves  to the above boundary  condition
(\ref{bcond}). In the course of the calculation of the coefficient  in
front of the divergent  wave  in the  ansatz (the scattering
amplitude) we  will use  the $Q$-function of the inner  operator
${\bf B}_0^+$. This parameter appears in the course of the solution of the
homogeneous equation for the adjoint operator as a map of symplectic
variables. The Krein's  $Q$-function  is an  abstract  analog  of  the
Weyl-Titchmarsh   $m$-function \cite{Titchmarsh}.
\par
\begin{lemma} \label{QF}
{The  $Q$-function for the operator ${\bf B}$ has  the
form:
\[
[e,\frac{I +  \lambda {\bf B}}{{\bf B} - \lambda I} e],
\]
where the  symplectic coordinates $\xi_{\pm}$ of  the  solution
$e_{\bar{\lambda}}$ of the adjoint homogeneous equation are
related by
     \[
\xi_- = - [e,\frac{I +  \lambda {\bf B}}{{\bf B} - \lambda I} e]
\xi_{+}.
\]
}
\end{lemma}
     {\it Proof}. Consider the $g$-orthogonal  projections $P_g = e ] [ e$
     and $ {\bf I}- P_g $  in the $g$-metric  onto the
one-dimensional deficiency subspace $N_i$ and  onto the
complementary subspace $E \ominus_g N_i$, respectively. Then  the solution of
the adjoint homogeneous equation $e_{\bar{\lambda}}$ may
be written as:
\[
e_{\bar{\lambda}} =\frac{{\bf B} + i I}{{\bf B} - \lambda I} e =
\frac{{\bf B}}{{\bf B} - i I}e + \frac{I}{{\bf B} - i I} e\,\,
[e,\frac{I + \lambda {\bf B}}{{\bf B} - \lambda I}\,\, e\,\,] +
     \]
\begin{equation}
\label{qfunc}  \frac{I}{{\bf B} - i I} ({\bf I}- P_g )\frac{I +
\lambda {\bf B}}{{\bf B} - \lambda I}e.
\end{equation}
The  last term in the  right-hand   side is an element $u_0$
from the domain of  the restricted operator ${\bf B}_0$, but the
first  two terms belong  to the defect  $M$ so  the  whole
linear combination may  be  written as
\[
\frac{{\bf B}+iI}{{\bf B} - \lambda I}e = w_+  -  Q (\lambda) w_- + u_0,
\]
where $Q (\lambda) =  [e,\frac{I +  \lambda {\bf B}}{{\bf B} -
\lambda I} e] $.

$\Box$

{\bf Remark} The Krein's $Q$-function  for a Hermitian operator in
a  space  with a {\it positive  metric} $g > 0$ is  a
dimensionless Nevanlinna-class function  with positive
imaginary part in the upper  half-plane ${\rm Im} \lambda > 0$ . If ${\bf
B}$ is a finite diagonal matrix diag$\displaystyle \left\{\lambda_s
\right\}_{s = 1}^{^N}$,\, $N = {\mbox{dim}E}$, and the  metric
tensor $g$ is trivial $g = I$, then
\[
Q(\lambda) = \sum_{s=1}^{N} \frac{1 + \lambda \lambda_s}
{\lambda_s - \lambda} |e^s|^2 ,
\]
where the $\lambda_s$ are non-negative  eigenvalues of ${\bf B}$ and
$|e^s|^2$ are
the squares of the moduli of the components $e^s$ of  the
deficiency  vector $e$ with respect to the (standard) basis of
eigenvectors  of ${\bf B}$ in $E$.
\par
Now  consider an  indefinite  metric tensor $g$  defined by
a diagonal  matrix, for  instance  $\left\{g_{ss}\right\}= \pm
1,\, s= 1,2,\dots N$, a real positive diagonal matrix ${\bf B} =
\left\{\lambda_{s}\right\},\,\, s = 1,2,\dots$ dim$E$ and  a
$g$-normalized  deficiency  vector  $e$ $([e,e]=1)$  with non-zero
components $e^s$ with respect  to the same standard  basis. Then
the corresponding $Q$-function will have a form
\begin{equation}
\displaystyle
     \label{q_indef} Q(\lambda) =
\sum_{s=1}^{N} \frac{1 + \lambda \lambda_s}{\lambda_s -\lambda}
g_{ss}|e^s|^2 = \sum_{s=1}^{N} \frac{1 + \lambda
\lambda_s}{\lambda_s -\lambda}P_s,
\end{equation}
with {\it real} coefficients $P_s = g_{ss}|e^s|^2$ and
   non-negative  eigenvalues  $\lambda_s$.
     The $Q$-function
(\ref{q_indef})  has poles of
first order at the eigenvalues $\lambda_s$ of  ${\bf B}$.
\par
     The previous lemma permits us to solve the adjoint
non-homogeneous equation and derive the Krein formula for the
resolvent of the self-adjoint (or at least self-adjoint in the $g$-metric)
extension ${\cal A}$ of the symmetric operator $l_0 \oplus {\bf
B}_0$ which is defined by the boundary condition (\ref{bcond}) (see the
simple derivation of the  Krein formula for  the  operator ${\bf
B}$ in Appendix A). The S-scattering matrix may be  derived  from it in
a rather standard way, see for instance \cite{Ad_Pav}. We
concentrate now on the direct derivation of an expression for the
scattering amplitudes and the construction of special
solutions---scattered waves---of
the homogeneous equation
\[
{\cal A} \Psi = \left(l_0^+ \oplus {\bf B}_0^+\right)\, \Psi = \lambda
   \Psi,
\]
which satisfy the boundary condition  (\ref{bcond}). The
scattered waves serve as eigenfunctions of the  self-adjoint
extension  $\cal A$ of the restricted operator  $l_0 \oplus{\bf
B}_0$ and may be found in the form
\[
\Psi = \left(
\begin{array}{c}
{\bf{\psi}}\\
\psi_E
\end{array}
\right),
\]
where the ansatz  for the  ``outer" component ${\bf {\psi}}$ of
the  scattered  wave in $R_3$ is a sum  of
     incoming and  outgoing waves:
\begin{equation}
{\bf{\psi}}_\lambda (x,\nu) =  e^{i \sqrt{\lambda}
(\nu,x)} +  T(\nu,\sqrt {\lambda})\frac{e^{i \sqrt{\lambda}
|x|}}{4\pi |x|},
\label{ET}
\end{equation}
with the  amplitude  $T(\nu,\sqrt {\lambda})$ in front of the  outgoing wave.
We note that this dimensionless amplitude $T$ is proportional to the standard
``physical" amplitude $f$ : $T=4\pi k_{0}f$. The
component $\psi_E$ of the scattered  wave in the  inner space $E$
is just proportional to the  limit  value of the solution
$e_{\bar{\lambda}}$ of the adjoint  homogeneous equation on the
real  axis of the  spectral parameter:
\[
\psi_E = T_E e_{\bar{\lambda}} = T_E \left[ w_+ -
Q(\lambda) w_- + u_0 \right].
\]
One  may  show, see  for instance \cite{Albeverio,Kurasov}, that
the boundary conditions (\ref{bcond}) formulated  for  elements of
the domain of  the extension of the  operator   $l_0 \oplus {\bf B}_0$
are valid for  the corresponding scattered  waves. The  symplectic
variables in the  above  ansatz for components  of the  scattered
wave in the outer and the inner  spaces  are:
\begin{equation}
B = \left( 1 + i\frac{\sqrt{\lambda}}{4 \pi} T \right),\,\, A
= T,\, \xi_- = - Q T_E ,\,\, \xi_+ = T_E.
\label{coeff}
\end{equation}
This gives the following equation for the amplitudes $T$ and $T_E$:
\begin{equation}
\label{ampeq}
\left(
\begin{array}{c}
   1 + i\frac{\sqrt{\lambda}}{4 \pi} T\\
Q T_E
\end{array}
\right) = \left(
\begin{array}{cc}
\gamma_{00} & \gamma_{01} \\
\gamma_{10} & \gamma_{11}
\end{array}
\right)
     \left(
\begin{array}{c}
    T \\
    T_E
\end{array}
\right).
\end{equation}
Solving this equation we obtain the following expressions for
these amplitudes.
\begin{theorem}
\label{samplitude}{ The  amplitudes  $T,$ and $T_E$
     for  real positive  values of the spectral variable $\lambda$ are equal to
\begin{equation}
\label{amplitudes} T(\sqrt{\lambda}) = \frac{1}{\gamma_{00}-
\frac{|\gamma_{01}|^2}{\gamma_{11} - Q} - i
\frac{\sqrt{\lambda}}{4 \pi }},\,\, T_E =
\frac{\gamma_{10}}{Q-\gamma_{11}}\,\, T,
\end{equation}
where
\[
   Q(\lambda)= \sum_{s=1}^{N} \frac{1+
\lambda \lambda_s}{\lambda_s - \lambda} |e^s|^2 g_{ss}.
\]
}
\end{theorem}
{\it Proof} may  be  obtained  from the  previous discussion.

{\bf Remark\,\,} Note that the  amplitude
\[
T (\sqrt{\lambda}) = \frac{1}{\gamma_{00}-
\frac{|\gamma_{01}|^2}{\gamma_{11} - Q} - i
\frac{\sqrt{\lambda}}{4 \pi }},
\]
contains essential  information on the spectral properties  of  the
extension ${\cal A}$. For example, the  solution of
the corresponding homogeneous equation
\[
{\cal A}\Psi = \lambda \Psi,
\]
\[
\Psi = \left(
\begin{array}{c}
\frac{e^{i
\sqrt{\lambda} |x|}}{4\pi |x|}\\
T_{_E} e_{\bar{\lambda}},
\end{array}
\right)
\]
yields  the  equation
\[
      \left(
     \begin{array}{c}
      i \frac{\sqrt{\lambda}}{4\pi}\\
      Q T_{_E},
      \end{array}
     \right) =
     \left(
\begin{array}{cc}
\gamma_{00}  & \gamma_{01} \\
\gamma_{10} & \gamma_{11}
\end{array}
\right) \,\, \left(
     \begin{array}{c}
      1 \\
     T_{_E}
      \end{array}
     \right)
\]
which has a  solution  $\Psi$ with  $T_{E}=
\frac{-\gamma_{10}}{\gamma_{11}- Q(\lambda)}$  if and only if the
{\it dispersion equation} is fulfilled:
\[
\frac{i\sqrt{\lambda}}{4\pi}= \gamma_{00} -
\frac{|\gamma_{01}|^2}{\gamma_{11}- Q(\lambda)}.
\]
The negative  roots $\lambda_{s} < 0$ of the  above  equation
   on the spectral  sheet ${\rm Im}\sqrt{\lambda}> 0 $ of  the
variable $\sqrt{\lambda}$ correspond to  the  square-integrable
first component of  the  solution $\Psi$, hence  they are  the
eigenvalues of the extended operator  ${\cal A}$.

\vskip2cm
\section{A Special class of Zero-range potentials
with Inner  structure}

In this section we study a special class
of  zero-range  models for  which the function $F(k)=k {\rm cot}\delta(k)$
   satisfies  an  analyticity  condition with respect to the wave number
   $k$ $([k]=cm^{-1})$.
   We will show that for models from  this  class the boundary parameters and
the deficiency vector
     in the finite-dimensional case
dim$\, E = N$ may be calculated directly. Moreover, the type of
indefinite metric in the inner space $E$ is also pre-determined  by
the above condition of analyticity.

   We  will also use the dimensional {\it
physical scattering amplitude}  $f(k)$ with orbital angular momentum
$l=0$, see \cite{Landau}, connected to the above dimensionless
amplitude $T(\sqrt{\lambda})$ in the ``outer'' space.
Using Eqs.(\ref{ET}) and (\ref{amplitudes}) we find this s-scattering
amplitude $f(k)$ as:
\begin{equation}
f (k)=\frac{1}{4\pi k_{0}} T(\sqrt{\lambda})= \left(4\pi
k_{0}\gamma_{00}-\frac{4\pi
k_{0}|\gamma_{01}|^{2}}{\gamma_{11}-Q(\lambda)}]-ik\right)^{-1}.
\label{s1}
\end{equation}
The scattering matrix $S(k)=\exp (2i\delta (k))$ where $\delta (k)$
is the scattering phase in the s-channel has the form:
\begin{equation}
S(k) =\frac{{\rm cot}\,\,\delta(k)+i}{{\rm
cot}\,\,\delta(k)-i}=1 + 2ik f (k), \label{s2}
\end{equation}
hence the  scattering amplitude $f(k)$ is:
\begin{equation}
f(k)=\frac{1}{k{\rm cot}\,\,\delta(k)-ik}.
\label{s3}
\end{equation}
Combining Eqs.(\ref{s1}), (\ref{s2}) and (\ref{s3}) one may find the
S-scattering
matrix in the form:
\begin{equation}
S(k)=  1 + \frac{2 i k}{4 \pi k_0
     [\gamma_{00}-\frac{|\gamma_{01}|^{2}}{\gamma_{11}-Q(\lambda)}]-ik} =
\frac{F(k) + ik}{F(k)- ik},
\label{A25}
\end{equation}
with
\begin{equation}
F(k)= 4 \pi k_0
[\gamma_{00}-\frac{|\gamma_{01}|^{2}}{\gamma_{11}-Q(\lambda)}].
\label{A26}
\end{equation}
    We may
rewrite the $Q$-function calculated in lemma 3.1 in terms  of the
wave number $k$, assuming that all eigenvalues
$\lambda_s=(k_{s}/k_{0})^{2}$ of  the inner Hamiltonian ${\bf B}$ are positive.
   Then it takes the following form
\begin{equation}
Q(\lambda)=\frac{1}{k_0^2}
\sum_{s=1}^{N}\frac{k_{0}^{4}+k_{s}^{2}k^{2}}{k_{s}^{2}-k^{2}}P_{s},
\label{A27}
\end{equation}
where the weights $P_s = g_{ss} |e^s|^2$ are not necessary positive!
Note that we have introduced the resonance  values $k_s > 0$ of the wave
number $k$ through the equation $\lambda_s=(k_{s}/k_{0})^{2}$, hence
$[k_{s}]=cm^{-1}$.
Using Eqs.~(\ref{A26}) and (\ref{A27}) we  obtain a
similar representation for the function $F(k)$
\[
F(k)\equiv k{\rm
cot}\delta(k)=4\pi\gamma_{00}k_{0}-
\frac{4\pi|\gamma_{01}|^{2}k_{0}}{\gamma_
{11}-
\frac{1}{k_0^2}\sum_{s=1}^{N}\frac{k_{0}^{4}+
k_{s}^{2}k^{2}}{k_{s}^{2}-
k^{2}}P_{s}}.
\]
\par
{\it We consider a  special type of  potential with inner
structure for  which  the  function $F(k)$ defined  by  the  above
formula is  an  {\it entire  function} of  the  variable $k$}.
\par
   The behavior of the function $F(k)$
defines  the scattering process and can be
represented by a Taylor expansion at the origin $k=0$:
\begin{equation}
F(k) = -\frac{1}{a} + \frac{r_0}{2 }k^2 +
\ldots=\sum_{n=0}^{\infty}g_{n}k^{2n}, \label{A28}
\end{equation}
where $a$ is  the {\it scattering length} and $r_0$ is the {\it effective
radius}.
\par
We may also impose  on the parameters of our model the requirement that the
scattering matrix satisfy  the following ``physically reasonable'' behaviour:
$S(k)\to 1$ when
$k^2 \to
\infty$. This condition is  fulfilled  if and
only if the denominator $\gamma_{11}- Q(\lambda)$ tends  to  zero
as $1/\lambda$ when $\lambda = (k/k_{0})^2 \to \infty$
which yields:
\begin{equation}
\gamma_{11}+\sum_{s=1}^{N}\frac{k^2_{s}}{k_0^2}P_{s}=0.
\label{A29}
\end{equation}
We will show that for the introduced class of zero-range
potentials the S-scattering matrix satisfies this ``physically reasonable''
condition for large energy automatically.
\begin{lemma}
\label{l1}{The function
\begin{equation}
F(k)=k{\rm cot}\delta(k)= 4\pi k_0\left[\gamma_{00}-
\frac{|\gamma_{01}|^{2}}{\gamma_{11}-
\frac{1}{k_0^2}\sum_{s=1}^{N}\frac{k_0^4 + k_s^2 k^2}{k_s^2 - k^2
}P_s }\right] \label{A30}
\end{equation}
is  an entire  function of $k$ (a  polynomial of degree $N$ in $k^2$)
\[
     F (k)\approx - 4 \pi |\gamma_{01}|^2 (-1)^N
     \frac{k^{^{2N}}}{\Lambda k_{0}^{2N-1}},
\,\,\, k \to \infty,
\]
with non-zero {\bf normalization constant} $\Lambda$, if and only
if the values of all weights $P_s$ are defined by the formula:
\begin{equation}
P_{s}= (-1)^{^{s}}
\frac{k_{0}^{2N+2}\Lambda}{(k_{0}^{4}+k_{s}^{4})\prod_{t (t\neq
s)}|k_{t}^{2}-k_{s}^{2}|}.
\label{A31}
\end{equation}
Here the  product $\prod_{t(t\neq s)}$ is spread over all indices
$t,\, 1 \leq t \leq N,\,$ except one, $t \neq s$. This analyticity requirement
automatically yields the  boundary parameter $\gamma_{11}$ as
\[
\gamma_{11}= - \sum_{s=1}^N \frac{k_s^2}{k_0^2}P_s.
\]
}
\end{lemma}
{\it Proof.}\,\,We may  consider the  denominator $D$ of the
$F$-function as  a function of the variable $\lambda = k^2 / k_0^2$
with  parameters  $\lambda_s = k^2_s/k_0^2$:
\begin{equation}
\label{fractions}
     D(\lambda)= \gamma_{11}- Q (\lambda)= \gamma_{11} +
\sum_{s=1}^{^N}\lambda_s P_s - \sum_{s=1}^{N}\frac{1+
\lambda_s^2}{\lambda_s - \lambda}P_s,
\end{equation}
where $N = \mbox{dim}E$. The analyticity of the fraction
$\frac{1}{D(\lambda)}$ on the  whole  plane $\lambda$ means that there
are no zeroes of the denominator in any compact domain of the complex $\lambda$
plane. Hence $D(\lambda)$, being a rational
function with $N$ simple poles must have  only  one  zero of
multiplicity $N$ at infinity and  hence both  $1/D(\lambda)$ and
$F(\lambda)$ are  polynomials of degree  $N$:
\[
     D(\lambda)\approx \frac{\Lambda}{\lambda^N} (-1)^N,\,\,\lambda \to \infty,
\]
where we introduce the  sign-factor $(-1)^N$ to give  the  final
formula a simple form. Then
the asymptotic of
the polynomial at  infinity is  defined  by  the highest order
coefficient :
\[
     \Lambda \neq 0,\, F (k)\approx - 4 \pi k_0 |\gamma_{01}|^2 (-1)^N
     \frac{\lambda^{N}}{\Lambda}.
\]
The constant $\Lambda$ plays an  essential role below. We  will  call
it  the {\it
normalization constant}.
\par On the other hand, expanding the fractions in the above
representation (\ref{fractions}) of $D$  in Laurent series we
obtain $D(\lambda)$ in another form:
\[
D(\lambda)= \gamma_{11}+ \sum_{s=1}^{^N}\lambda_s P_s +
\frac{1}{\lambda}\sum_{s=1}^{^N}( 1+ \lambda^2_s) P_s
+\frac{1}{\lambda^{2}}\sum_{s=1}^{N}(1+\lambda^{2}_{s})\lambda_{s}P_{s}
\]
\begin{equation}
\label{eqs}  +\dots +\frac{1}{\lambda^{^N}}\sum_{s=1}^{^N}(
1+ \lambda^2_s) \lambda_s^{N-1}P_s + o(\lambda^{^{-N}})\approx
(-1)^N \frac{\Lambda}{\lambda^N},\,\, \lambda \to \infty.
\end{equation}
This means that all coefficients in front of the powers
$\lambda^{-l}$ ($0\leq l < N $) are  equal to  zero and the
coefficient in front of $\lambda^{^{-N}}$ is  equal to the
normalization constant. Thus the variables $q_s =
(1+\lambda_s^2)P_{s}$ ($s=1,2,\dots,N$) fulfill the linear system
obtained by comparison of the coefficients  in front of the  powers
$\frac{1}{\lambda^l}$ ($l=1,2,\dots, N$) on the  left and  right hand
sides of  the last equation (\ref{eqs}). This system has a
positive Vandermonde determinant $W = W(\lambda_1,\, \lambda_2,\,
\lambda_3,\, \dots \lambda_N)$ if the  parameters $\lambda_s$ are
arranged in order of increasing $ \lambda_1 < \lambda_2<
\lambda_3< \dots \lambda_N$. The  solution of it may be found as a
ratio of the determinant and minors  $W_s = W(\lambda_1,\,
\lambda_2,\,\ldots, \lambda_{s-1},\lambda_{s+1}\dots
\lambda_N)> 0$:
\begin{equation}
\label{solution}
     (1+\lambda_s^2)P_s = (-1)^{^s}\frac{W_s}{W}
\Lambda= \frac{(-1)^{^s}}{\prod_{t(t\neq s)}|\lambda_s -
\lambda_t|}\Lambda,\,\, s=1,2,\dots N.
\end{equation}
The announced result is obtained from the last formula  by
inserting wave numbers $k_s,\, k$ instead of $\lambda_s,\,\lambda$.

$\Box$

{\bf Corollary 1} We  may  assume (see theorem 4.1), that the  metric tensor is
defined  as  $g_{ss}= (-1)^{^{s}}$ sgn$\Lambda$. Then the squares
of the components $|e^s|^2$ of the  deficiency vector $e$ are given by
\begin{equation}
\label{defvec}
     |e^s|^2 = \frac{1}{1 +
\lambda_s^2}\frac{|\Lambda|}{\prod_{t(t\neq s)}|\lambda_s -
\lambda_t|}= \frac{|\Lambda|k_0^{2(N+1)}}{(k_0^4 + k_s^4
)\prod_{t(t\neq s)}|k^2_s - k^2_t|}.
\end{equation}
\par
{\bf Remark}. The  basic  formula for the  $Q$-function was
derived under the assumption that the  deficiency vector $e$ is
non-degenerate with respect to  the indefinite metric form and
$[e,e]= 1$. In the ``generic case"
\begin{equation}
\label{nondeg}
     \sum_{s=1}^N\frac{1}{1 +
\lambda_s^2}\,\,\,\frac{(-1)^{^{s}}}{\prod_{t(t\neq s)}|\lambda_t
-\lambda_s|} \neq 0,
\end{equation}
so we  may  choose the  normalization constant as
\[
\Lambda =\frac{1}{\sum_{s=1}^N\frac{1}{1 +
\lambda_s^2}\,\,\,\frac{(-1)^{^{s}}}{\prod_{t(t\neq s)}|\lambda_t
-\lambda_s|}}=
\frac{1}{\sum_{s=1}^N \frac{k_0^{^{2(N+1)}}(-1)^{^{s}}}{(k_s^4 +
k_0^4 ) \prod_{t(t\neq s)}|k^2_t - k^2_s|}},
\]
and  use the  normalized deficiency vector in the  formula for the
$Q$-function. {\it We assume that the non-degeneracy
condition (\ref{nondeg})  is  fulfilled}.

\begin{theorem}
\label{t1} {The necessary condition for the analyticity
of the function $F(k)=k{\rm cot}\delta(k)$ on a whole complex
plane of $k$ for zero-range potentials with inner structure
is the alternation of the diagonal components of the metric tensor $g$:
\begin{equation}
g_{ss}=(-1)^{^{s}}{\rm sign}\Lambda, \label{A38}
\end{equation}
     where the  order of the components
($s=1,2,\ldots$) is defined by the ordering of the parameters
$k_{s}$~: $k_{1}^{2}<k_{2}^{2}<\ldots$. The  only parameters which
remain free in the  model are the scattering length, the effective
radius, and the eigenvalues of  the  inner
Hamiltonian which  commutes  with the  metric  tensor. The
boundary  parameters $\gamma_{00},|\gamma_{01}|$  are defined
as  functions  of the  scattering length  $a$, effective  radius
$r_0$, and $k_{s}^{2}~~(s=1,2,\ldots,N.)$:
\[
a =\left(-4 \pi \gamma_{00}
k_0+\frac{4\pi|\gamma_{01}|^{2}}{k_{0}^{2N-1}\Lambda}\prod_{s=1}^{N}k_{s}^{2
}\right)^{-1} ,\,\,
r_{0}=\frac{8 \pi |\gamma_{01}|^2}{\Lambda \, k_0^{^{2N-1}}}
\left(\prod_{t=1}^{N}k_{t}^{2}\right)\sum_{s=1}^{N}k_{s}^{-2}.
\]
In particular, if  the  effective  radius  is  positive, $r_0 > 0$,
     then  $\Lambda > 0$ and  $g_{ss} = \left( -1 \right)^{s}$.
All
other essential parameters of  the model: the  moduli  of the components of
the  deficiency  vector and  the  the  inner  boundary  parameter
$\gamma_{11}$ are  uniquely  defined from  the condition of
analyticity  of  the  $F$-function on the  whole $k$ plane.}
\end{theorem}
{\it Proof.} Using Eq.(\ref{A31}) and our condition
$k_{1}^{2}<k_{2}^{2}<\ldots$ one see that
\begin{equation}
{\rm sign}P_{s}=(-1)^{^{s}}{\rm sign}\Lambda. \label{A39}
\end{equation}
Combining this equation  with our definition $P_{s}=g_{ss}|e^{s}|^{2}$ we
find the
necessary condition for the metric tensor $g$ given by
Eq.(\ref{A38}).
The explicit form of the function $F(k)$ follows from
Eqs.(\ref{A30}) and (\ref{A31}):
\begin{equation}
F(k)=k{\rm cot}\delta(k)= \varepsilon  - \gamma
\prod_{s=1}^{N}(k_{s}^{2}-k^{2}),
\label{A40}
\end{equation}
where we have introduced  the  notations
\begin{equation}
\varepsilon =4\pi\gamma_{00}k_{0},~~\gamma=\frac{4\pi |
\gamma_{01}|^2}{k^{2N-1}_{0}\Lambda}. \label{A41}
\end{equation}
In particular, using (\ref{A40}), (\ref{A41}) and definition
(\ref{A28}) we find the scattering length $a$ and effective radius
$r_{0}$ :
\begin{equation}
a=\left(-\varepsilon +\gamma\prod_{s=1}^{N}k_{s}^{2}\right)^{-1},
r_{0}=2\gamma \sum_{s=1}^{N}\prod_{t(t\neq s)}k_{t}^{2}.
\label{A42}
\end{equation}
This completes the theorem.

$\Box$

{\bf Corollary 2}  It is remarkable that, using (\ref{A25}) and
(\ref{A40}),
we  may  write
the scattering  matrix $S$ in terms of the parameters
$k_{s}$, $\varepsilon$ and $\gamma$ as:
\begin{equation}
S(k)= 1+\frac{2ik}{\varepsilon -ik-\gamma\prod_{s=1}^{N}(k_{s}^{2}-k^{2})}.
\label{A43}
\end{equation}
The total scattering cross-section is $\sigma(k)=4\pi|f(k)|^{2}$ or
can be written in
explicit form (see next setion) using  equations (\ref{s3}) and (\ref{A40}) .
This expression for the S-scattering matrix describes the
resonance scattering of  particles with resonances  defined  by  the
spectral  properties of the inner  Hamiltonian (again see the next section).
\vskip1cm

\vskip1cm
\section{Resonance s-scattering on Zero-range potentials
with Inner Structure }

In this last section we consider the main physical properties of the
resonance scattering problem on zero-range potentials with inner
structure. This
problem has application mainly to low-energy neutron scattering by
nuclei.

First we will show that zero-range
potentials with inner structure, as defined above, fulfil our condition for
{\it generalized zero-range potentials} (\ref{i15}), ie. the special
boundary condition at the origin ${\bf r}=0$. Due to the fact that
the condition
(\ref{i15}) follows from the expansion (\ref{i16}) we will
consider Eq.~(\ref{i16}) which can be rewritten in dimensionless form
(\ref{fermivar}) as:
\begin{equation}
u(x,k) = \frac{A(k)}{4\pi |x|}+B(k)+o(1),
\label{l1}
\end{equation}
where
\begin{equation}
A(k) = 4\pi k_{0}C{\rm sin}\delta(k),~~~B(k)=kC{\rm cos}\delta(k).
\label{l2}
\end{equation}
Hence the special boundary condition (\ref{i15}) is equivalent to the
equation:
\begin{equation}
\frac{4\pi B(k)}{A(k)} = \frac{k}{k_{0}}{\rm cot}\delta(k).
\label{l3}
\end{equation}
On the other hand, as follows from Eqs.~(\ref{coeff}) and
(\ref{amplitudes}), the
left side of Eq.~(\ref{l3}) for zero-range potentials with inner
structure is:
\begin{equation}
\frac{4\pi B(k)}{A(k)} =
\frac{4\pi}{A(k)}+i\sqrt{\lambda}=4\pi\left(\gamma_{00}-
\frac{|\gamma_{01}|^{2}}{\gamma_{11}-Q(\lambda)}\right).
\label{l4}
\end{equation}
Using Eq.~(\ref{A26}) we find that the right side of Eq.~(\ref{l4}) is
$k_{0}^{-1}F(k)=(k/k_{0}){\rm cot}\delta(k)$, hence equations
(\ref{l3}) and (\ref{i15}) are satisfied for zero-range potentials with inner
structure.
Thus our definition (\ref{i15}) is correct for the zero-range
potentials derived in this paper.

Let us show that our resonance scattering model depends only the
scattering length $a$, effective radius $r_{0}$, and the spectrum
$k_{s}$ ($s=1,2,\ldots,N$) of the inner
Hamiltonian. It is convenient to define the typical wave-number $k_{0}$
and a dimensionless parameter $\alpha$  by:
\begin{equation}
\label{50}
k_0: =- \frac{1}{4 \pi \gamma_{00} a_0},~~~\alpha: = - \frac{4 \pi
|\gamma_{01}|^2}{\Lambda} \left( 4
\pi \gamma_{00} \right)^{2N-1},
\end{equation}
where $a_{0}$ is a new parameter with dimension ($[a_{0}]=cm$).
Then  the function $F(k)= k \cot \delta (k)$ for our zero-range
potential is:
\begin{equation}
\label{51} F(k) =  -\frac{1}{a_{0}}- \frac{\alpha}{a_{0}} \prod_{s =1}^{N}
\left(a_0^2 k_s^2 - a_0^2 k^2\right).
\end{equation}
Defining  the  scattering length $a$ from the  equation
$F(0)=-a^{-1}$, we  obtain:
\begin{equation}
\label{52}
a = \frac{a_{0}}{1 + \alpha \prod_{s =1}^{N} a_0^2 k_s^2}.
\end{equation}
Hence the parameter $a_0$ is the  scattering length for
the zero-range  potential without inner structure, i.e. when $\alpha
=0$. Evidently the  above  equation (\ref{52}) is the renormalization
equation since it expresses the  renormalized scattering length $a$ via the
non-renormalized scattering length $a_{0}$, taking into account resonance
scattering.

Thus the function $F(k)$ given by Eq.~(\ref{51}) depends on $N+2$
parameters: $a_0, \alpha$ and $ k_s,\, s = 1,2,\dots N$. The  effective
radius may
be  found  from (\ref{51}) as
\begin{equation}
\label{53}
\frac{r_0}{2} = \alpha  a_0 \sum_{n=1}^N \prod_{s(s\neq n)}^{N} a_0^2 k_s^2 .
\end{equation}
Consequently, the  non-renormalized scattering length $a_0$, the
dimensionless parameter  $\alpha$ and the spectrum
$k_{s}$ ($s=1,2,\ldots,N$) of the
inner Hamiltonian define the scattering
length $a$ and effective  radius $r_{0}$ of the model. Vice  versa, we  may
choose  the renormalized  scattering  length $a$, the  effective
radius $r_{0}$, and the spectrum
$k_{s}$ ($s=1,2,\ldots,N$) of the
inner Hamiltonian as  independent
parameters, and  define  $a_0$ and
$\alpha$ by Eqs.(\ref{52}) and (\ref{53}) but  the former option is more
useful and  will be used below.
\par
The total cross-section  for  spherically-symmetric scattering
is calculated  as
\begin{equation}
\sigma (k)= 4\pi |f(k)|^2 = \frac{4\pi }{|F(k)- ik|^2},
\label{cross}
\end{equation}
which  implies,  due  to  (\ref{51}), the explicit formula:
\begin{equation}
\label{54} \sigma (k)= \frac{4 \pi a_0^2}{1 + a_0^2k^2 +
2\alpha\prod_{s=1}^N \left(a_0^2 k_s^2 - a_0^2 k^2\right) +
\alpha^2 \prod_{s=1}^N \left(a_0^2 k_s^2 - a_0^2 k^2\right)^2}.
\end{equation}
The latter equation exhibits clear  resonance  properties: for
$k=k_s$ the  products in the  denominator vanish and  the  whole
cross-section is  reduced  to
\[
\sigma (k_s)= \frac{4 \pi a_0^2}{1 + a_0^2k_s^2},~~~ s= 1,2,\dots
N,
\]
which coincides with the Wiegner formula $\sigma (k)=\frac{4\pi
a_{0}^{2}}{1+a_{0}^{2}k^{2}}$ for $k=k_{s}$.
\par
    Consider  now the  analytic structure of  the scattering matrix
    (\ref{A25})
    \[
S(k)= \frac{F(k) + ik}{F(k) - ik}
    \]
in the complex $k$ plane. Then the equation for the poles of the scattering
matrix is:
\[
F(k)-ik=0.
\]
One may  see that all  solutions  of  this  equation may  be written
as  $k = i\kappa$, where $\kappa$ are  the zeroes of another  polynomial
with real  coefficients:
\begin{equation}
\label{55}
\alpha\prod_{s =1}^{N} \left(a_0^2 k_s^2 + a_0^2
\kappa^2\right) - a_{0}\kappa +1 = 0.
\end{equation}
In the simplest  case $N=0$ one  may  see  from (\ref{51}), (\ref{52}),
and (\ref{55}):
\[
a_0 = a,\, \kappa = \frac{1}{a}, \, F(k) = - \frac{1}{a} =-\kappa,
\]
which is in full  agreement  with the Fermi model, see \cite{Fermi} and  the
Introduction   above. From Eq.~(\ref{54}) it follows that the
integral cross-section in this case is given by the  classical Wiegner formula:
$\sigma (k) = \frac{4 \pi a^2}{1 + a^2 k^2}$, where $a=a_{0}$.
\par
    For $N=1$ one  may  find  from (\ref{55}):
\begin{equation}
\label{56}
\frac{r_0}{2}\kappa^2 - \kappa + \frac{1}{a}=0
    \end{equation}
with
\[
\frac{1}{a}= \frac{1}{a_0} + \alpha a_0 k_1^2,\,\, \frac{r_0}{2}=
\alpha a_0.
\]
Hence Eq.~(\ref{51}) in this case yields:
\[
F(k)=-\frac{1}{a}+\frac{r_{0}}{2}k^{2}.
\]
These results and Eq.~(\ref{54}) for $N=1$  coincide with well  known
results for  low-energy
resonance  scattering, see \cite{Landau} and Eqs.~(\ref{i9}) and
(\ref{i10}).
\par
In the case $N=2$ one  may  show from equations (\ref{55}),
(\ref{52}), and (\ref{53})
that:
\begin{equation}
\label{57}
    \frac{r_0}{2\left( k_1^2 + k_2^2\right)} \kappa^4 +
\frac{r_0}{2}\kappa^2 - \kappa + \frac{1}{a}=0,
\end{equation}
    with
    \[
\frac{1}{a}= \frac{1}{a_0} + \alpha a_0^3 k_1^2
k_2^2,~~~\frac{r_0}{2}= \alpha a_0^3 \left( k_1^2 + k_2^2\right).
    \]
In this case the total cross-section is given by Eq.~(\ref{54}) for
$N=2$.
In the general case $N>2$ it follows from Eq.~(\ref{55}) that some of
the poles of the
S-scattering  matrix are
situated on the  real  axis $\kappa$ (imaginary axis
$k=i\kappa=p+iq$) and the other
sit  at complex-conjugated  points  $\kappa_s=q_{s}-ip_{s},
\bar{\kappa}_s=q_{s}+ip_{s}$, i.e.
symmetric with respect  to  the  real  axis in $k$-plane, because
all coefficients of the algebraic  equation (\ref{55}) are real. The
zeroes  of  the  S-scattering  matrix are  defined  by the adjoint
equation $F(k)+ ik=0$, hence if a pole of the S-scattering matrix is
situated at the point $k_{s}=p_{s}+iq_{s}$ then the appropriate zero is
situated at the complex-conjugated point $\bar{k}=p_{s}-iq_{s}$.
Note that  these are  universal properties  of  the
S-scattering  matrix  and  are  a consequence of
unitarity and  causality, but we have derived  them using only
Eq.~(\ref{55}). The  imaginary poles $k_{s}=iq_{s}$  may lie on both
the positive
or negative  imaginary
semi-axes ($q_{s}>0$ or $q_{s}<0$), but all other poles (when $p_{s}\neq
0$) should lie in the lower half-plane. This  property  is  connected  with
general  spectral
properties  of  selfadjoint  operators in spaces  with  indefinite
metric, see  for  instance \cite{Krein-Langer,Shondin1, Shondin2}.
Thus the properties of  the  function  $F(k)$ given by Eq.~(\ref{51})
guarantee the correct analytic structure of  the S-scattering  matrix in the
complex plane of the wave-number $k$.
    \par
    For  our  model  the  poles $k_n = p_n + i q_n$ of  the
S-scattering  matrix  may  be
     classified  according  to  their  positions  using the following scheme:

     1. If $p_{n}=0$ then for $q_{n}>0$ the pole corresponds to a bound
     state with the energy $E_n = -\frac{\hbar^2}{2\mu}q_n^2$ and for
     $q_{n}<0$ the pole corresponds to a virtual state with
     energy $E_n = -\frac{\hbar^2}{2\mu}q_n^2$.

     2. Poles  with  $ p_n> 0,\,\,  q_n< 0 $ correspond  to
     metastable states with complex energy $E_n = E_n' - i
     \frac{\Gamma_n}{2}$, where:
     \[
     E'_n = \frac{\hbar^2}{2\mu}\left[ (p_n)^2 - (q_n)^2\right],\,\,
     \Gamma_n = - \frac{\hbar^2}{2\mu} p_n q_n > 0
     \]
     and the rate of decay of the metastable state is
     $w_n=\frac{\Gamma_n}{\hbar}$. Here $E_n' >0 $ if  $p_n > |q_n|$.

     3. Poles  with $ p_n< 0,\,\,  q_n< 0 $ correspond  to
     resonance trapping. These  poles  and
     the  corresponding poles  of  the  metastable states are
symmetrically situated
     with respect  to  the  imaginary  axis.

     No  other  poles  are  present  in  our  zero-range  model since all
     poles  may  be  found  as  zeroes  of  a    polynomial with
      real  coefficients.

      In conclusion we point out the main problems which should be
  examined in the future in connection with the model of zero-range potentials
with inner structure.
  Besides pure resonance scattering which we have considered in this
  paper, scattering theory of nucleons by nuclei must
  include:

  Potential scattering connected with consideration of the
  interaction around the nucleus surface and with the Coulomb
  interaction in the case of proton scattering by nuclei.
  The spin-orbit interaction describing polarization effects.
  The generalization of the model for scattering with an
  arbitrary quantum number $l$.

   \section{Acnowledgements}
The authors  are  grateful  to  Professor H. Langer for
discussions of the structure and properties of  the positive
invariant subspace of  self-adjoint operators in Pontryagin
space and  to  doctors  Y. Shondin and  P. Kurasov for attracting
our attention to the paper \cite{Fewster} where the  question of
fitting  parameters was  considered  for a class of zero-range
potentials {\it without inner structure}. Following  advice  from
them the  authors  included the Appendix which contains a
simple derivation of the  classical  Krein Formula  for  resolvents
of  self-adjoint  extensions  in  Pontryagin spaces.
B.P. is grateful to the Japanese Foundation JSPS which supported
his stay at the Solid State Computer Laboratory of Aizu University,
Japan, where a part of this paper was written, and  to Professor V. Ryzhii
    for  useful advice.

\section{Appendix A: Derivation of the Krein formula
     with symplectic  coordinates}

In this Appendix, using the notations $B_{0}={\bf B_{0}}$ and $B={\bf B}$,
we give a simple  derivation of  the  classical
Krein formula for  the  $g$-symmetric operator $B_0$ in the
Pontryagin space $E$. This derivation follows the  pattern of
\cite{Extensions} where  similar  formulae  for  operators in
spaces with  positive metric is  presented and may be easily
modified for extensions of the operator $l_0 \oplus B_0$
in Pontryagin space.
\par
Consider a  $g$-hermitian  operator $B$ in  the
Pontryagin space $E$ with  an indefinite  dot-product $[.,.]=
\langle g .,.\rangle$. Assume that  the  operators  $B \pm i I$
are  invertible, and  ${\cal E}\subset E$ is  a finite-dimensional
subspace which will play the role of  the deficiency subspace $M_i$
for  the  reduced operator  $B_0$  defined  on $D_0 = \frac{I}{B -
iI }E,\,\, M_i = \left\{\tilde{x} : \left[ \left( B-iI \right) D_0 ,
\tilde{x}
\right]=0 \right\} $. The deficiency subspace at the point $-i$
may be  obtained from ${\cal E}$ via a g-unitary transformation
$M_{-i}= \frac{B + iI}{B - iI} {\cal E}$. If  the operator  $B_0$
is  densely  defined then  the subspaces form a positive  angle.
If the  operator $B_0$ is  not  densely defined, (for  instance,
if $B$ is  bounded) then  {\it we   assume} that the  angle between
$M_{\pm i} $ is  positive. In both cases one may define  an
adjoint  of the formally self-adjoint operator, $B_0^+$, at least on the
defect $M = M_i + M_{-i}$. In fact the extension of the reduced
operator $B_0$   must be constructed on the defect and hence we  may
derive the Krein formula below only on the defect.
\par
The subspace ${\cal E}$ is  mapped into the  defect by  the
transformations
\[
\displaystyle
\begin{array}{ccc}
e&\longrightarrow &\frac{e + \frac{B + iI}{B - iI}e}{2} =
\frac{B}{B-iI}e := w_+,\\
e&\longrightarrow &\frac{e - \frac{B + iI}{B - iI}e}{2i} =
\frac{-1}{B-iI}e := w_-,
\end{array}
\]
where
\[
B_0^{^{+}} w_+ = w_-,\,\, B_0^{^{-}} w_- = - w_+.
\]
Then the  solution  $u = u_0 + \frac{B}{B-iI}\xi_+ -
\frac{I}{B-iI}\xi_-$of  the non-homogeneous  adjoint equation
\[
B^{^{+}}u = B u_0 - \frac{I}{B-iI}\xi_+ - \frac{B}{B-iI}\xi_- =
\lambda u + f
\]
may be found from the condition:
\[
\left( B- \lambda I\right)u_0 - \frac{I + \lambda B}{B - iI}\xi_+
+ \frac{B-\lambda I}{B - iI}\xi_- = f.
\]
We require that the  symplectic coordinates $\xi_{\pm}$ satisfy the
following boundary
condition in terms of the $g$-Hermitian  operator $\Gamma$:
\[
\xi_-  + \Gamma \xi_- = 0,
\]
and find an  explicit formula for  the resolvent of the of  the
corresponding  extension $B_{\Gamma}$. Assuming that the operator
$B-\lambda I$ is invertible we may multiply the above equation by
$\frac{B-iI}{B - \lambda}$, thus obtaining
\[
\left( B - i I\right)u_0 = \frac{B-iI}{B -\lambda I}f + \frac{I
+\lambda B}{B-\lambda I}\xi_+  + \xi_-.
\]
Substituting here  $\xi_-$ from the  above boundary  condition and
projecting orthogonally (in the
Pontryagin space) onto the  subspace ${\cal E}$ we obtain an
important connection
between the symplectic coordinate $\xi_{+}$ of the  solution $u$ and  $f$:
\begin{equation}
\label{KF} \xi_+ = \frac{I}{\Gamma - P_{_{\cal E}}\frac{I +
\lambda B}{B - \lambda I}P_{_{\cal E}}}\,\,\,\frac{B-iI}{B
-\lambda I}f .
\end{equation}
This  connection permits us to  calculate the  resolvent of  the
extension:
\[
u = \left[B_{\Gamma} - \lambda I\right]^{^{-1}}f =
\]
\[
\label{K1} \frac{I + \lambda B}{B - \lambda I}\,\,\,\frac{I}{B -
iI} \xi_+ - \frac{I}{B - iI}\,\, \Gamma \xi_+  + \frac{I}{ B -
\lambda I} f  + \frac{B}{B - iI} \xi_+ \frac{I}{B - iI}\,\, \Gamma
\xi_+ =
\]
\[
\label{K2} \frac{I}{ B - \lambda I} f  + \frac{I + B^2}{\left( B-
\lambda I \right)\left(B - iI \right)} \,\,\, \xi_+ =
\]
\begin{equation}
\label{KF} \frac{I}{ B - \lambda I} f  + \frac{B + iI}{B - \lambda
I }\,\,\, \frac{I}{\Gamma - P_{_{\cal E}}\frac{I + \lambda B}{B -
\lambda I}P_{_{\cal E}}}\,\,\,\frac{B-iI}{B -\lambda I}f.
\end{equation}
The  last  expression gives  the  Krein  formula for  the
     resolvent of  the  extension  $B_{_{\Gamma}}$ of the reduced operator.

\section{Appendix B: Singular $\delta$-potentials}

We consider here potentials which are often used as an approximation
of, for example, a hard sphere potential or a 3-dimensional deep square well
potential \cite{PW98} when the atom-atom interactions are effectively weak
and dominated by elastic s-wave scattering:
\begin{equation}
V({\bf r}-{\bf r}') = U_{0} \delta({\bf r}-{\bf r}'),
\label{B1}
\end{equation}
where
\begin{equation}
U_{0}=\frac{2\pi a
\hbar^{2}}{\mu},~~~\frac{1}{\mu}=\frac{1}{m_{1}}+\frac{1}{m_{2}}.
\label{B2}
\end{equation}
Here $\mu$ is the reduced mass ($\mu=m/2$ when the scattering particles
are identical) and $a$ is the s-wave scattering length. The sign of
the parameter $a$ ($a>0$ or $a<0$ for repulsive and attractive interactions
respectively) depends sensitively on the precise details of the
interatomic potential. This potential is considered a good
approximation when three-body processes can be neglected and
hence we need take only binary collisions into account. Note
that expression (\ref{B2}) for $U_{0}$ is found by the Born approximation
in the case of low-energy ($k\rightarrow 0$) scattering (though this
approximation is is not valid here, see below). Moreover one can suppose
that the potential (\ref{B1}) yields the total cross-section as
$\sigma=4\pi a^{2}$ in the limit $k\rightarrow 0$ when the scattering
particles are different and $\sigma=8\pi a^{2}$ for identical
particles (Bose statistics). Evidently this potential is
pointlike and hence, in some sense, one may consider it as a simple
example of a zero-range potential. For $U_{0}<0$ this
potential represents a deep well, see the Introduction of this paper.
Nevertheless we show below that the potential (\ref{B1}) is not the
correct approximation for atom-atom collisions in both cases $a>0$
and $a<0$ and hence cannot represent some kind of zero-range
potential. Actually this fact is well known. However, we examine it in
detail as it demonstrates some non-trivial aspects of the theory of
zero-range potentials and because this $\delta$-singular
potential is still very popular \cite{PW98}.

We may consider the Dirac $\delta$-function in (\ref{B1}) as a
$\delta$-sequence $\delta_{\varepsilon}({\bf  r}-{\bf r}')$ given by:
\begin{equation}
\delta_{\varepsilon}({\bf  r}-{\bf r}') = \left\{
\begin{array}{ccc}
\frac{3}{4\pi \varepsilon^{3}r_{0}^{3}}   ,\,&\mbox{if}& |{\bf
r}-{\bf r}'| \leq \varepsilon r_{0}\\
0 ,\,&\mbox{if}& |{\bf r}-{\bf r}'|> \varepsilon r_{0}
\end{array} \right.  .
\label{B3}
\end{equation}
Then the potential (\ref{B1}) is  $V(r)=V_{0}$ if $r\leq \varepsilon
r_{0}$, and $V(0)=0$ if $r>\varepsilon r_{0}$, where
\begin{equation}
V_{0}=\frac{3a\hbar^{2}}{2\mu \varepsilon^{3}r_{0}^{3}}.
\label{B4}
\end{equation}

In case 1 when $U_{0}>0~~(a>0)$ and $k\rightarrow 0$ the total
cross-section is \cite{Landau}:
\begin{equation}
\sigma = 4\pi \varepsilon^{2}r_{0}^{2}\left(\frac{{\rm
tanh}(\varepsilon q_{0}r_{0})}{\varepsilon q_{0}r_{0}}-1\right)^{2},
\label{B5}
\end{equation}
where $q_{0}=\sqrt{2\mu V_{0}}\hbar^{-1}\sim \varepsilon^{-3/2}$.

For $\varepsilon \rightarrow 0$ we have $\sigma\sim
\varepsilon^{2}$ and hence the total cross-section $\sigma$ for the
singular potential (\ref{B1}) is zero which means that no scattering
occurs in this case!

We note that the conventional derivation of the Gross-Pitaevskii
equation is based on the potential (\ref{B1}). Fortunately this
equation may be found \cite{Note} without using the $\delta$-approximation, see
also \cite{Krug}.

In case 2 when $U_{0}<0~~(a<0)$ the potential (\ref{B1})
represents an infinitely deep well: $V(r)=-|V_{0}|$ if $r\leq \varepsilon
r_{0}$, and $V(0)=0$ if $r>\varepsilon r_{0}$. Then the total
cross-section for $k\rightarrow 0$ is:
\begin{equation}
\sigma = 4\pi \varepsilon^{2}r_{0}^{2}\left(\frac{{\rm
tang}(\varepsilon q_{0}r_{0})}{\varepsilon q_{0}r_{0}}-1\right)^{2},
\label{B6}
\end{equation}
with $q_{0}=\sqrt{2\mu |V_{0}|}\hbar^{-1}\sim \varepsilon^{-3/2}$.
The limit $\varepsilon \rightarrow 0$ does not exist in this equation which
means that the potential (\ref{B1}) is not correct in the case
$U_{0}<0~~(a<0)$ at all.

Thus the singular potential (\ref{B1}) is not a proper approximation
for scattering processes in both cases 1 and 2. Actually the
correct approximation of the hard sphere potential (case 1) is given in the
papers \cite{HY} and \cite{Lee}, see also \cite{Huang}, and the correct
approximation for the deep
well (case 2) was developed by Fermi \cite{Fermi}, see also the
Introduction.

\end{document}